\documentclass[12pt,preprint]{aastex}

\begin{document}

\title{Testing Disk Instability Models for Giant Planet Formation}

\author{Alan P.~Boss}
\affil{Department of Terrestrial Magnetism, Carnegie Institution of
Washington, 5241 Broad Branch Road, NW, Washington, DC 20015-1305}
\authoremail{boss@dtm.ciw.edu}

\begin{abstract}

 Disk instability is an attractive yet controversial means for
the rapid formation of giant planets in our solar system and
elsewhere. Recent concerns regarding the first adiabatic exponent
of molecular hydrogen gas are addressed and shown not to lead to
spurious clump formation in the author's disk instability
models. A number of disk instability models have been calculated
in order to further test the robustness of the mechanism,
exploring the effects of changing the pressure equation of
state, the vertical temperature profile, and other parameters affecting
the temperature distribution. Possible reasons for differences in
results obtained by other workers are discussed. Disk instability
remains as a plausible formation mechanism for giant planets.

\end{abstract}


\section{Introduction}

 The disk instability mechanism for giant planet formation is
based on the formation and survival of self-gravitating clumps
of gas and dust in a marginally gravitationally unstable
protoplanetary disk (Boss 1997; reviewed by Durisen et al. 2007).
In order for a disk instability to succeed, the disk must be
able to cool its midplane as the clumps form, allowing them
to continue to contract to higher densities, and the clumps
must be able to survive indefinitely in the face of Keplerian
shear, tidal forces, and internal thermal pressure. Considering
the complicated physical processes involved during the time
evolution of a three dimensional disk instability, it is perhaps
not surprising that the theoretical basis for the disk
instability hypothesis remains unclear even after a decade of
work on the subject.

 The Indiana University (IU) group has been active in studying disk
instabilities, and has generally found that disk instabilities
are unable to lead to the formation of self-gravitating, dense
clumps that could go on to form gas giant protoplanets (e.g., Pickett et
al. 2000; Mej\'ia 2004; Cai et al. 2006a,b; Boley et al. 2006, 2007a,b).
On the other hand, the Washington-Zurich group (e.g., Mayer et al.
2002, 2004, 2007) has presented models that support the hypothesis
that disk instabilities can lead to the formation of long-lived
giant gaseous protoplanets. We present here
several new calculations that attempt to understand the reasons
for some of these different outcomes regarding disk instability.

\section{Energy Equation of State}

 Boley et al. (2007a) pointed out that uncertainties about the
ortho/para ratio of molecular hydrogen at low temperatures might
lead to differences in the outcome of disk instability models.
They suggested that a mixture intermediate between pure parahydrogen
and a 3:1 ortho/para ratio might be the most appropriate choice.
Boley et al. (2007a) also noted that discontinuities in the
specific internal energy equation for molecular hydrogen could
lead to artificially low values of the first adiabatic exponent
for a simple perfect gas ($\Gamma_1 = \gamma = 1 + R_g/(\mu c_V$),
where $R_g$ is the gas constant, $\mu$ is the mean molecular weight,
and $c_V$ is the specific heat at constant volume; Cox \& Giuli 1968).
Values of $\gamma \le 4/3$ can lead to dynamical instabilities
(either expansion or contraction) away from a configuration of
hydrostatic equilibrium (Cox \& Giuli 1968). Boley et al. (2007a)
suggested that $\gamma \le 4/3$ could artificially lead to clump
formation.

 Boley et al. (2007a) stated that the energy equation of
state (EOS) used in disk instability models by Boss (2001, 2002b, 2005)
contained a discontinuity that might be responsible for artificially
lowering $\gamma$ below 4/3, leading to spurious fragmentation.
Boley et al. (2007a) based this assertion on the equations of
state described by Boss (1984). However, all of the hydrodynamical
models run by the present author since Boss (1989) have been based on a
different energy equation than that reported by Boss (1984), as a
result of a direct comparison with the equation of state routines
employed by Werner Tscharnuter. The models of Boss (1989) revised
the energy equation treatment to include an interpolation
between temperatures of 100 K and 200 K, but this revision was
not explicitly stated in the Boss (1989) paper as it seemed
insignificant at the time.

 Using the notation of Boss (1984), the specific internal
energy for molecular hydrogen for temperatures $T < 100 K$ is
taken to be $E^*_{H_2} = 3/2 \ R_g T/\mu$, while for temperatures
between $100 K$ and $600 K$ it is $E^*_{H_2} = 5/2 \ R_g T/\mu$.
For intermediate temperatures, $100 K < T < 200 K$, the
internal energy is interpolated according to the equation
$E^*_{H_2} = 3/2 \ R_g T/\mu \ [1 + 2/3 \ (T - 100)/100]$.
The Erratum by Boley et al. (2007b) used this revised EOS.

 Boley et al. (2007a,b) showed that for a pure parahydrogen or 3:1
ortho:para mix, $\gamma$ decreases significantly at $\sim 100 K$,
but does not fall below 4/3. Figure 1 depicts the behavior of
$\gamma$ calculated with the revised Boss equation of state and shows
that $\gamma$ drops below 4/3 for $135 K < T < 200 K$. The
densest clumps found in Boss's disk instability models generally have
maximum temperatures below $135 K$, and even lower mean temperatures.
Boss (2005) presented a disk instability model with the highest spatial
resolution computed to date, and found that the densest clump that formed had
a maximum temperature of $120 K$ and a mean temperature of $94 K$.
Boss (2006c) found that clumps formed in disk instabilities around M dwarf
protostars had maximum temperatures less than $100 K$. Boss's (2002a)
Table 1 showed maximum clump temperatures of $115 K$ to $126 K$ for a range
of models with varied opacities.

 Perhaps the most important point raised by Boley et al. (2007a, b) is
that $\gamma$ is likely to decrease significantly around $100 K$,
and this softening of the pressure EOS will enhance the formation of
dense clumps, as happens in locally isothermal disks with an
effective $\gamma = 1$. In fact, Figure 1 in Boley et al. (2007b) shows
that the Boss $\gamma$ is higher than that of either of the preferred
hydrogen mixtures for $T < 100 K$, implying that clump formation is
suppressed somewhat in the Boss models as a result.

\section{Pressure Equation of State}

 Several new models have explored using the same pressure EOS as is
used in the IU group models (e.g., Cai et al. 2006a,b; Boley et al.
2006): the gas pressure $p$ is given by $(\gamma -1) \rho E$,
where $\gamma = 5/3$, $\rho$ is the gas density, and $E$ is the
specific internal energy of the gas. Three models were run
with this pressure EOS, starting from the same initial disk
model as model HR in Boss (2001) -- a $0.091 M_\odot$ disk orbiting
a $1 M_\odot$ protostar with an outer disk radius of 20 AU.
The calculations were made with the same three dimensional, gravitational,
radiative hydrodynamics code as in Boss (2001) and in all subsequent
Boss disk instability calculations (see these earlier papers for more
details about the calculational techniques and initial conditions). The
models had $N_\phi = 256$ and $N_{Ylm} = 32$, though, compared to
$N_\phi = 512$ and $N_{Ylm} = 48$ for model HR (Boss 2001).

 The three new models varied the choice for the critical disk
density ($\rho_{cr} = 10^{-13}$, $10^{-12}$, or $10^{-11}$ g cm$^{-3}$)
below which the disk temperature was forced back to its initial
value (typically $40 K$ in the outer disk). This artifice was employed
in Boss (2001) and subsequent models in order to maintain a
reasonably large time step when low density regions develop in the disk
that are undergoing decompressional cooling.

 In all three models, spiral arms and transient clumps form
within 200 yrs of evolution, similar to the behavior with the
usual (Boss 1984) pressure EOS (e.g., Boss 2001). Relatively high
density clumps form, with maximum densities similar to those in a calculation
with the same spatial resolution but the usual pressure EOS.
Evidently using the $\gamma = 5/3$ EOS does not alter the results in a
significant way because the Boss EOS also has $\gamma = 5/3$ for
$T < 100 K$, the regime where clumps form. The choice of
$\rho_{cr}$ makes little difference as well, as was found by
Boss (2006b) for disk instability models in binary star systems.
These results suggest that the reason for differing outcomes must
be sought elsewhere (see Discussion).

\section{Varied Disk Temperature Parameters}

 We now present a set of four models varying several of the parameters
that could affect the temperature distribution in the disk models.
Table 1 summarizes the four models, which are all variations
on model HR of Boss (2001). However, these models all had the same
spatial resolution ($N_\phi = 512$) and number of terms in the
spherical harmonic expansion for the gravitational potential
($N_{Ylm} = 48$) as model HR in Boss (2001). The parameters that
were varied included: the temperature of the thermal bath (i.e., the
envelope temperature $T_e$), the critical density in the disk below
which the temperature was reset to the initial disk temperature at that
orbital radius (as in the previous section), the critical density in the
envelope below which the gas was assumed to be at a temperature equal
to the envelope temperature $T_e$ (for grid points at least 8 degrees
above the disk midplane), and finally, whether the temperature was forced
to decline monotonically (Boss 2002a) with vertical height inside the disk
(mono) or not (free). The former constraint errs on the side of artificially
cooling the disk by removing local temperature maxima in the vertical
direction. All models started at a time of 322 yrs of evolution in model
HR, and continued for at least another 17 yrs of evolution ($\sim 1$
clump orbital period).

 The results of these variations on the standard assumptions
are shown in Figures 2 and 3. The variation that produced the largest
deviation from the standard assumptions (model H in Figure 2) was relaxing
the constraint on the monotonic vertical (more precisely, in the
$\theta$ angle) decline of temperatures within the disk
(model TZ in Figure 3), though even this model led to the formation
of well-defined clumps that were no more than a factor of two less
dense than in model H. Models T and TE led to evolutions that were
very similar to that of model H and so are not shown. The models
show that these three variations in the details of how the disk
thermodynamics is treated in the Boss models are not particularly
significant for the outcome of a disk instability, presumably because
of the thermal bath assumption.

\section{Discussion}

 There are a number of possible reasons for different outcomes
compared to other groups:

{\it Spatial resolution} -- Clump formation is strongly enhanced
as the spatial resolution in the critical azimuthal direction is
increased from $N_\phi = 64$ to 512 (Boss 2000). Boss (2005)
presented a model with $N_\phi = 1024$ and a locally refined radial
grid (equivalent to a calculation with over $8 \times 10^6$ grid
points) that implied that in the continuum limit, the outcome
of a disk instability is the formation of dense, self-gravitating
clumps. Cai et al. (2006a,b) calculated models with $N_\phi = 128$,
increasing $N_\phi$ to 512 for two models only after those
models had entered a phase of evolution when nonaxisymmetry
was no longer growing. Boley et al. (2006) similarly calculated models
with $N_\phi = 128$, increasing $N_\phi$ for some models to 512
for the earliest phase of evolution, leading to the formation of
dense clumps at the intersections of spiral structures. The clumps
disappeared in a fraction of an orbital period. Clumps typically last
no more than an orbital period in even the highest spatial
resolution models of Boss (2005). Boss (2005) thus used virtual
protoplanets to allow the orbital evolution of these dense clumps
to be followed further than is possible with even a high spatial
resolution calculation with a fixed Eulerian grid code.

{\it Gravitational potential solver} -- The Boss models use a spherical
harmonic ($Y_{lm}(\theta, \phi)$) expansion to solve Poisson's equation
for the gravitational potential, with the accuracy of the
resulting gravitational potential being strongly dependent
on the number of terms ($N_{Ylm}$) carried along in the expansion.
As we have seen, model HR in Boss (2001) used $N_{Ylm} = 48$.
Boss (2000, 2001) found that increasing $N_{Ylm}$ led to the
formation of significantly denser clumps. Boss (2005) further
explored the effects of using an enhanced gravitational potential
solver by replacing some of the mass in the densest regions
of a clump with a point mass at the center of the relevant grid cell,
finding that this led to even better defined, higher density
clumps. In comparison, the IU group uses a direct solution of
Poisson's equation, with a boundary potential employing
terms up to $l = m = 10$ (Pickett et al. 2000), implying
a limited ability to depict small-scale gravitational forces in a
strongly nonaxisymmetric disk. Mej\'ia (2004) considered Fourier
analysis of her disk models for $m \le 6$, while Boley et al.
(2006) considered $m \le 63$, finding increasingly little power
in modes with $m > 10$, possibly consistent with the cutoff at
$l = m = 10$ in their boundary potential.

{\it Artificial viscosity} -- Pickett \& Durisen (2007) found
that the inclusion of certain artificial viscosity (AV) terms could
enhance the survival of clumps formed in a disk instability,
and suggested that AV could thus explain the long-lived clumps
found in SPH calculations by Mayer et al. (2002, 2004, 2007). Pickett \&
Durisen (2007) further noted that even calculations without any
explicit AV (e.g., the Boss models) should be considered suspect,
given the intrinsic numerical viscosity of any numerical code.
Boss (2006b) showed that with a large amount of explicit AV,
clump formation is suppressed, as was also found by Pickett \&
Durisen (2007). However, the level of intrinsic numerical
viscosity in Boss code models with $N_\phi = 256$ appears to be
equivalent to an $\alpha$-viscosity of $\sim 10^{-4}$ or smaller
(Boss 2004), a level that appears to be negligible in
comparison to typical explicit AV levels. Considering that
the virtual protoplanet (VP) models of Boss (2005) had $N_\phi = 256$,
the continued survival of the VP for at least 30 orbital periods
in these models is not likely to have been affected by the intrinsic
numerical viscosity of the Boss code.

{\it Stellar irradiation} -- Mej\'ia (2004) considered
the effects of stellar irradiation on the surface of the disk
as a means of heating the disk surface and thereby possibly
suppressing clump formation. The Boss models assume that
the disk is immersed in a thermal bath appropriate for
backscattering from infalling dust grains in the protostellar
envelope, with an envelope temperature appropriate for a
protostar that is not undergoing an FU Orionis outburst
(Chick \& Cassen 1997). While the Boss models thus
do not include the effects of direct irradiation by the
central protostar, the dynamical evolution of a three
dimensional disk leads to strongly variable vertical
oscillations and structures (Mej\'ia 2004; Boley \& Durisen 2006;
Jang-Condell \& Boss 2007) that are not considered in simple
theoretical models of flared accretion disks. The highly
corrugated inner disk surfaces (stretching at least 29 degrees above the
disk midplane; Jang-Condell \& Boss 2007) will shield the outer disk
surfaces from the central protostar, eliminating this source
of heating for much of the outer disk.

{\it Radiative transfer} -- Boley et al. (2006) have calculated
disk instability models using the flux-limited diffusion
approximation (FLDA) along with a detailed treatment of
the transition from the optically thick disk to the optically
thin atmosphere of the disk. They suggest using a plane-parallel (one
dimensional) atmosphere as a test case. Boss (2001) investigated the effects
of using the FLDA instead of the standard diffusion approximation
(DA) coupled with a thermal bath for low optical depths, but
did not find any significant differences. Myhill \& Boss (1993)
showed the results of the fully three dimensional, standard nonisothermal
test case for protostellar collapse, calculated with their two independent
Eddington approximation (EA) codes, finding good agreement.
Whitehouse \& Bate (2006) found that their FLDA models of
the standard nonisothermal collapse problem led to temperature
profiles similar to those found by Myhill \& Boss (1993), though with
appreciably hotter gas temperatures where the optical depth was
$\sim$ 2/3. They attribute this difference to the FLDA retarding
the loss of radiation in these layers compared to the EA.
Given that the choice of the flux limiter can have an effect
on the outcome (Bodenheimer et al. 1990), it is unclear whether
any particular implementation of the FLDA is superior to the EA.
The standard Boss models have used the DA coupled with a thermal
bath to force the DA models to mimic an EA calculation.

{\it Numerical heating} --  In calculations by the IU group,
nonaxisymmetric perturbations tend to grow rapidly for a certain
period of time and then begin to damp out (e.g., Cai et al. 2006a,b).
In the Boley et al. (2006) calculations, the disk starts out
with a mass of 0.07 $M_\odot$ and a radius of 40 AU, but then
expands outward to a radius of $\sim$ 80 AU, leading to the formation
of rings inside 20 AU and a gravitationally stable region outside
20AU with spiral arms that do not fragment. The latter behavior
is roughly consistent with the models by Boss (2003), who studied
disks extending from 10 AU to 30 AU, and found fragments
to form at 20 AU but not at 30 AU. Similarly, Boss (2006a)
studied disks extending from 100 AU to 200 AU, and found no
evidence for fragmentation. Thus on scales larger than $\sim 20$ AU,
the results of Boley et al. (2006) and Boss (2003, 2006a) are
in basic agreement. The disagreement arises about what happens
in the inner disks. Fragmentation typically occurs
at 8 to 10 AU in the Boss models, whereas the
inner disk rings do not fragment in the Boley et al. (2006)
models. Boley et al. (2006) state that their inner disk is stable
to ring fragmentation because of ``numerical heating'' at distances
out to $\sim 7$ AU. This non-physical heating appears to have affected
the models of Cai et al. (2006a,b) as well as those of Boley et al. (2006),
making the IU results for inner disks difficult to accept.

\section{Conclusions}

 While there are a number of potential reasons for the differences
in disk instability models calculated by the IU group and the
present author, at this time the major sources of these differences
would appear to be some combination of several effects, namely
spatial resolution, gravitational potential solver accuracy,
and numerical heating in the inner disk of the IU models. Handling of
the boundary conditions for the disk's radiative energy losses is
another possibility that is still under investigation by the author,
though the models presented here suggest that this may not be a dominant
effect. Given the current state of knowledge, and the new results presented
herein, it appears that reports of the death of the disk instability
model for giant planet formation have been greatly exaggerated.

 I thank Aaron Boley, Kai Cai, and Megan Pickett for working with me to
understand the differences between their group's results and my results,
and the referees for their remarks.
This research was supported in part by NASA Planetary Geology and Geophysics
grant NNG05GH30G and is contributed in part to NASA Astrobiology Institute
grant NCC2-1056. The calculations were performed on the Carnegie Alpha
Cluster,
the purchase of which was partially supported by NSF Major Research
Instrumentation grant MRI-9976645.

\clearpage
\begin{deluxetable}{ccccc}
\tablecaption{Models with varied disk temperature parameters.\label{tbl-1}}

\tablehead{\colhead{model} &
\colhead{$T_e$ (K) } &
\colhead{$\rho_{cr}$ (disk) } &
\colhead{$\rho_{cr}$ (envelope) } &
\colhead{$T(\theta)$ } }

\startdata

H  & 50K  &  $10^{-11}$  & $10^{-11}$ & mono \\

T  & 100K &  $10^{-13}$  & $10^{-11}$ & mono \\

TZ & 50K  &  $10^{-13}$  & $10^{-11}$ & free \\

TE & 50K  &  $10^{-13}$  & $10^{-13}$ & mono \\

\enddata
\end{deluxetable}
\clearpage


\begin{figure}
\vspace{-2.0in}
\plotone{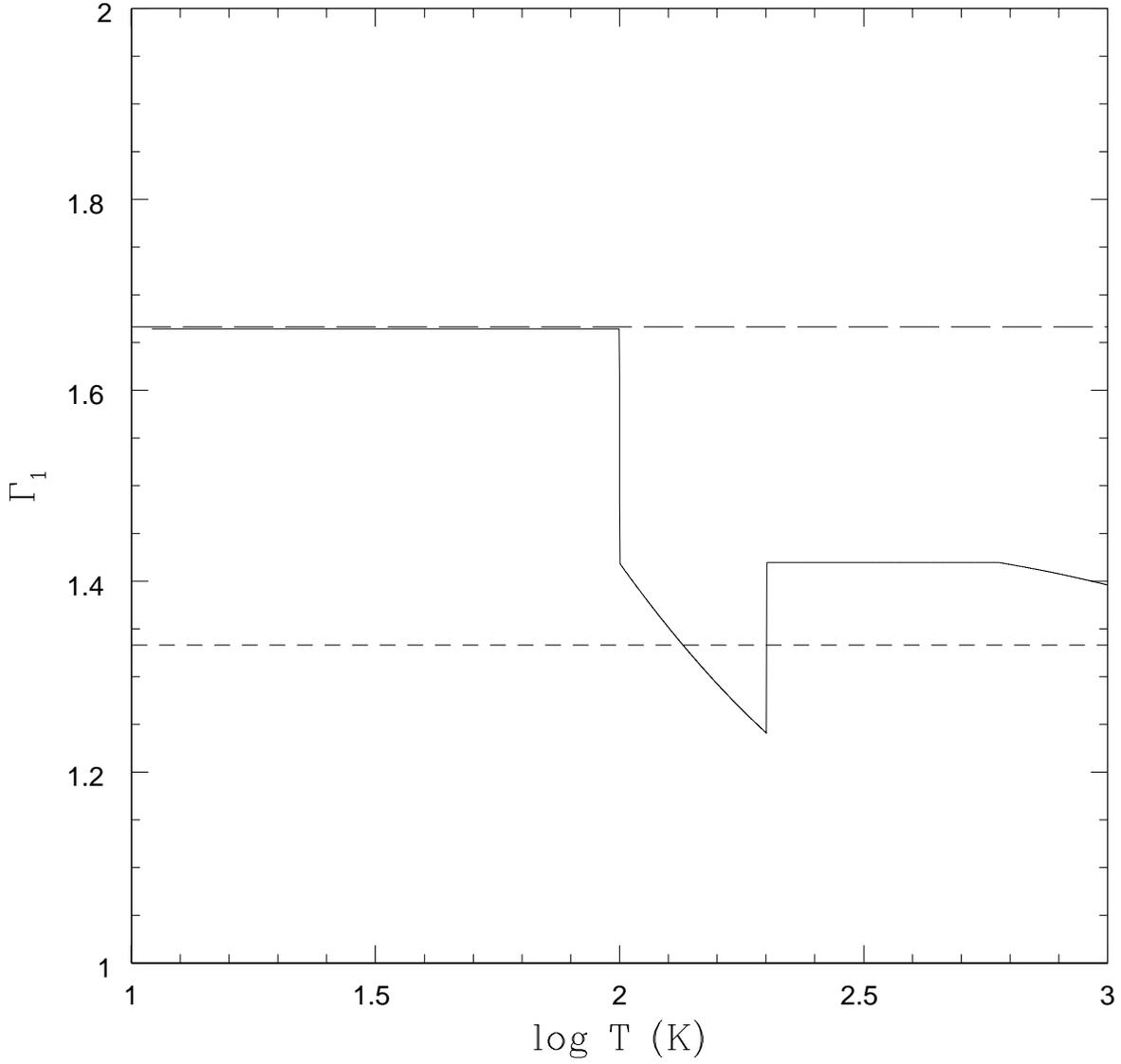}
\caption{Adiabatic exponent $\Gamma_1 = \gamma$ used in Boss (2001)
and in all subsequent Boss disk instability models (solid line).
The short-dashed line shows $\gamma = 4/3$, while the long-dashed line
shows $\gamma = 5/3$, as used by Cai et al. (2006a,b) and Boley et al.
(2006).}
\end{figure}

\clearpage

\begin{figure}
\vspace{-2.0in}
\plotone{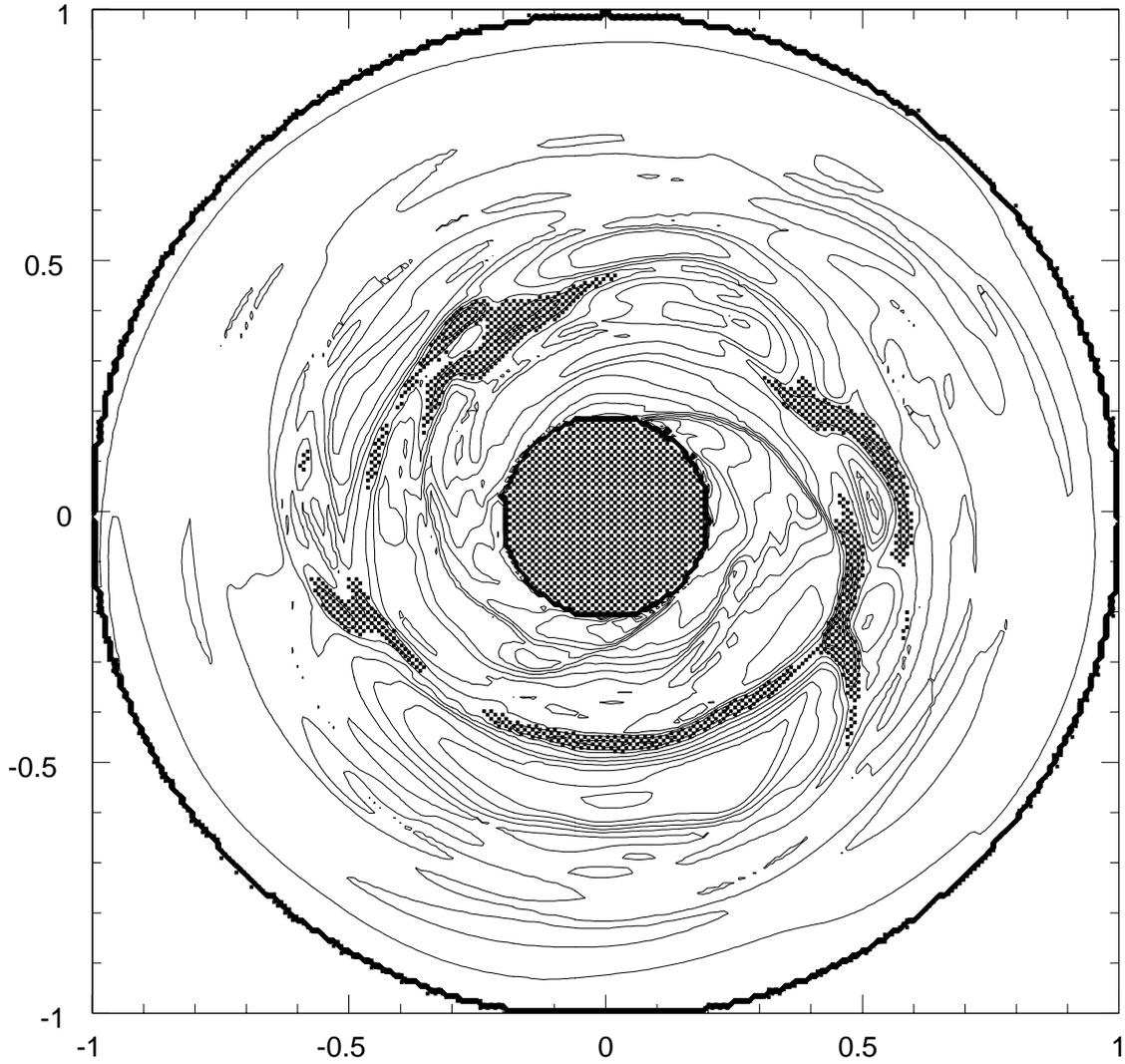}
\caption{Equatorial density contours for model H after 339 yrs of evolution.
The disk has an outer radius of 20 AU and an inner radius of 4 AU.
Hashed regions denote clumps and spiral arms with densities higher than
$10^{-10}$ g cm$^{-3}$. Density contours represent factors of two
change in density.}
\end{figure}

\clearpage

\begin{figure}
\vspace{-2.0in}
\plotone{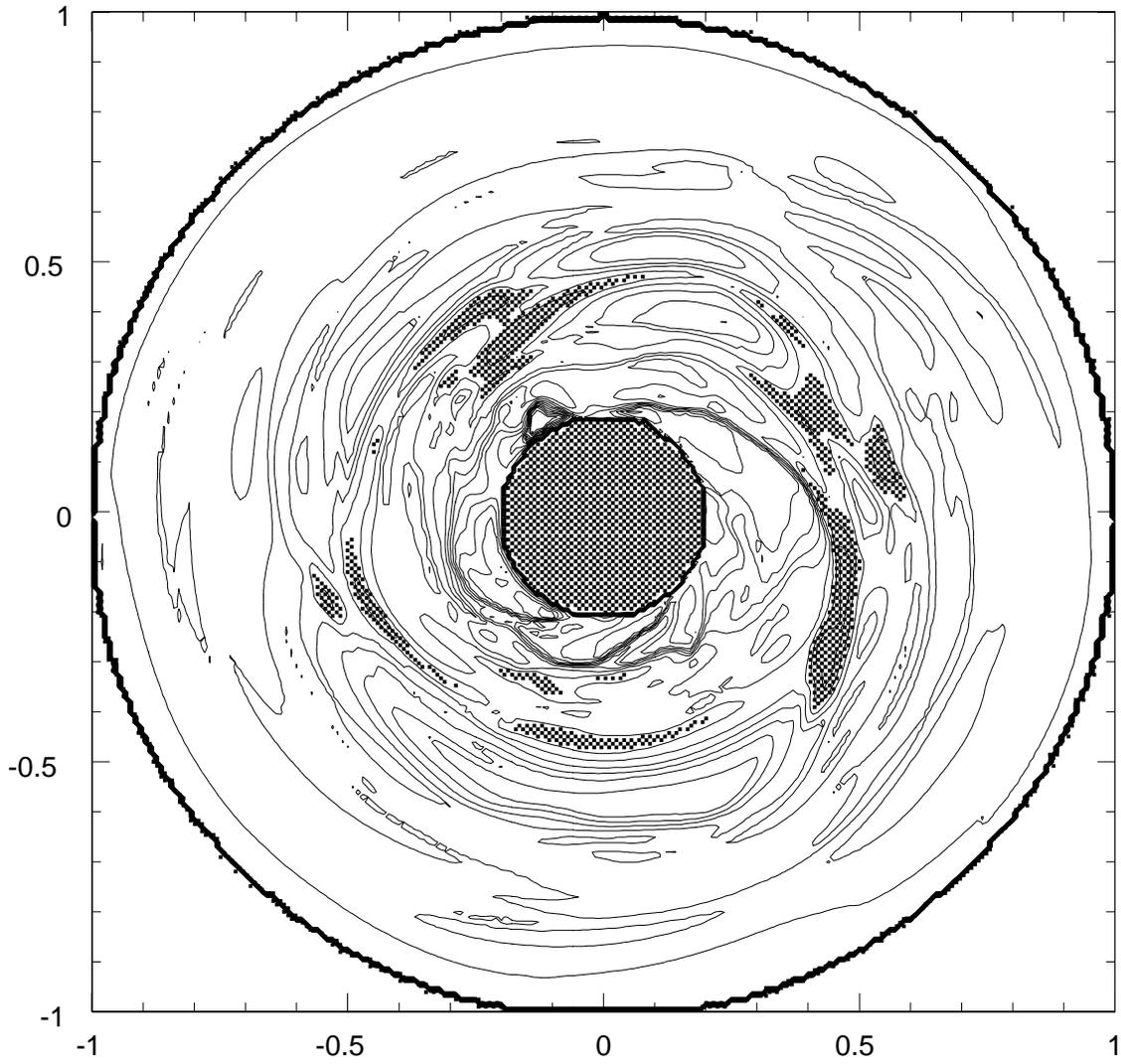}
\caption{Same as Figure 2, but for model TZ.}
\end{figure}

\end{document}